\documentclass[12pt]{article}
\usepackage{amsmath}
\usepackage{amssymb}
\usepackage{fancyhdr}
\renewcommand{\maketitle}{
    \begin{center}
        {\bf Higher-Order Kinetic Term for Controlling Photon Mass in\\
Off-Shell Electrodynamics}
        \vskip .3 true cm
      \normalsize
        Martin Land \\
        \vskip .3 true cm
        Department of Computer Science \\
        Hadassah College \\
        P. O. Box 1114, Jerusalem 91010, Israel \\
        email: martin@multinet.net.il
      \end{center}
      \vskip .5 true cm
}
\begin{document}

\title{}
\author{}
\maketitle

\begin{abstract}
In relativistic classical and quantum mechanics with Poincare-invariant
parameter, particle worldlines are traced out by the evolution of spacetime
events. The formulation of a covariant canonical framework for the evolving
events leads to a dynamical theory in which mass conservation is demoted
from {\it a priori} constraint to the status of conserved Noether current
for a certain class of interactions. In pre-Maxwell electrodynamics --- the
local gauge theory associated with this framework --- events induce five
local off-shell fields, which mediate interactions between instantaneous
events, not between the worldlines which represent entire particle
histories. The fifth field, required to compensate for dependence of gauge
transformations on the evolution parameter, enables the exchange of mass
between particles and fields. In the equilibrium limit, these pre-Maxwell
fields are pushed onto the zero-mass shell, but during interactions there is
no mechanism regulating the mass that photons may acquire, even when event
trajectories evolve far into the spacelike region. This feature of the
off-shell formalism requires the application of some {\it ad hoc} mechanism
for controlling the photon mass in two opposite physical domains: the low
energy motion of a charged event in classical Coulomb scattering, and the
renormalization of off-shell quantum electrodynamics. In this paper, we
discuss a nonlocal, higher derivative correction to the photon kinetic term,
which provides regulation of the photon mass in a manner which preserves the
gauge invariance and Poincare covariance of the original theory. We
demonstrate that the inclusion of this term is equivalent to an earlier solution to the
classical Coulomb problem, and that the resulting quantum field theory is
renormalized.
\end{abstract}

\baselineskip 7mm \parindent=0cm \parskip=10pt

\section{Introduction}

\subsection{Pre-Maxwell Theory}

The Stueckelberg equation \cite{Stueckelberg, H-P} 
\begin{equation}
i\partial _{\tau }\psi (x,\tau )=\frac{1}{2M}p^{\mu }p_{\mu }\psi (x,\tau )\
,  
\label{eqn:1}
\end{equation}
describing a free particle in the quantum mechanics of Horwitz and Piron \cite{H-P}, is
rendered locally gauge invariant \cite{saad} under gauge transformations 
\begin{equation}
\psi (x,\tau )\longrightarrow \left[ \exp ie_{0}\Lambda (x,\tau )\right] \
\psi (x,\tau )  
\label{eqn:2}
\end{equation}
when extended to the form 
\begin{equation}
(i\partial _{\tau }+e_{0}a_{5})\ \psi (x,\tau )=\frac{1}{2M}(p^{\mu
}-e_{0}a^{\mu })(p_{\mu }-e_{0}a_{\mu })\ \psi (x,\tau )\ ,  
\label{eqn:3}
\end{equation}
with gauge compensation fields which transform as 
\begin{equation}
a_{\mu }(x,\tau )\rightarrow a_{\mu }(x,\tau )+\partial _{\mu }\Lambda
(x,\tau )\qquad a_{5}(x,\tau )\rightarrow a_{5}(x,\tau )+\partial _{\tau
}\Lambda (x,\tau )\ .  
\label{eqn:4}
\end{equation}
Adopting the conventions 
\begin{equation}
x^{5}=\tau \qquad \qquad \partial _{5}=\partial _{\tau }\qquad \qquad
\lambda ,\mu ,\nu =0,1,2,3\qquad \qquad \alpha ,\beta ,\gamma =0,1,2,3,5
\label{eqn:designations}
\end{equation}
reframes equation (\ref{eqn:4}) as a five-dimensional symmetry
transformation 
\begin{equation}
a_{\alpha }(x,\tau )\rightarrow a_{\alpha }(x,\tau )+\partial _{\alpha
}\Lambda (x,\tau )  
\label{gauge}
\end{equation}
inducing --- from equation (\ref{eqn:3}) --- a five-dimensional conserved
current 
\begin{equation}
\partial _{\alpha }j^{\alpha }=0  
\label{current}
\end{equation}
where 
\begin{equation}
j^{5}=\Bigl|\psi (x,\tau )\Bigr|^{2}\qquad j^{\mu }=\frac{-i}{2M}\Bigl\{\psi
^{*}(\partial ^{\mu }-ie_{0}a^{\mu })\psi -\psi (\partial ^{\mu
}+ie_{0}a^{\mu })\psi ^{*}\Bigr\}\ .  
\label{eqn:6}
\end{equation}
Equations (\ref{current}) and (\ref{eqn:6}) permit the interpretation of $
\Bigl|\psi (x,\tau )\Bigr|^{2}$ as the probability density at $\tau $ of
finding the event at $x$. Since $\partial _{\mu }j^{\mu }=-\partial _{\tau
}j^{5}\neq 0$, we may not identify $j^{\mu }$ as the source current in
Maxwell's equations. However, under the boundary conditions $
j^{5}\rightarrow 0$, pointwise, as $\tau \rightarrow \pm \infty $,
integration of (\ref{current}) over $\tau $, leads to $\partial _{\mu
}J^{\mu }=0$, where 
\begin{equation}
J^{\mu }(x)=\int_{-\infty }^{\infty }d\tau \ j^{\mu }(x,\tau )\ .
\label{eqn:8}
\end{equation}
This integration has been called concatenation \cite{concat} and links the
event current $j^{\mu }$ with the particle current $J^{\mu }$ defined on the
entire worldline. The quantum mechanical potential theory with $a_{\mu }=0$
and $-e_{0}\phi =V(\sqrt{x^{\mu }x_{\mu }})$ has been solved for the
standard bound state \cite{bound} and scattering \cite{scattering} problems.
Some justification is provided for the pre-Maxwell theory by the covariant
treatment of the Zeeman \cite{zeeman} and Stark \cite{stark} effects for the
hydrogen-like bound state: the Horwitz-Piron theory reproduces the expected
line splitting phenomenology, only when the pre-Maxwell scalar current is
included.

The associated classical mechanics is obtained by transforming the
Hamiltonian found from (\ref{eqn:3}) to a classical Lagrangian \cite{emlf},
and including the gauge invariant kinetic term for the fields proposed by
Sa'ad, et.\ al.\ \cite{saad}: 
\begin{equation}
L=\frac{1}{2}M\dot{x}^{\mu }\dot{x}_{\mu }+e_{0}\dot{x}^{\alpha }a_{\alpha }-
\frac{\lambda }{4}f^{\alpha \beta }f_{\alpha \beta }\ .  
\label{eqn:14}
\end{equation}
The field strength tensor
\begin{equation}
f_{\alpha \beta }=\partial _{\alpha }a_{\beta }-\partial _{\beta }a_{\alpha }
\label{eqn:13}
\end{equation}
is chosen --- as in standard Maxwell theory --- to be first order in the
fields and manifestly gauge invariant. Variation of (\ref{eqn:14}) with
respect to $x^{\mu }$ leads to the classical Lorentz force \cite{emlf} 
\begin{equation}
M\;\ddot{x}^{\mu }=e_{0}f_{\;\;\alpha }^{\mu }(x,\tau )\,\dot{x}^{\alpha
}\qquad \qquad \frac{d}{d\tau }(-\frac{1}{2}M\dot{x}^{2})=e_{0}\ f_{5\alpha }
\dot{x}^{\alpha }\ .  
\label{eqn:12}
\end{equation}
and exchange of mass between particles and fields may be seen in the second
of equation (\ref{eqn:12}); the total mass-energy-momentum of the events and
fields is, however, conserved \cite{emlf}. Since particle mass is not
separately conserved, pair annihilation is classically permitted.

In formally raising the index $\beta =5$ in $f^{\mu 5}=\partial ^{\mu
}a^{5}-\partial ^{5}a^{\mu }$, Sa'ad et.\ al.\ argue that the action
suggests a higher symmetry containing O(3,1) as a subgroup, that is, either
O(4,1) or O(3,2). They wrote the metric for the field as 
\begin{equation}
g^{\alpha \beta }={\rm diag}(-1,1,1,1,\sigma )\ ,  
\label{eqn:15}
\end{equation}
where $\sigma =\pm 1$, depending on the higher symmetry. Variation of (\ref
{eqn:14}) with respect to $a_{\alpha }$ yields 
\begin{equation}
\partial _{\beta }f^{\alpha \beta }=\frac{e_{0}}{\lambda}j^{\alpha }=e\ j^{\alpha }\qquad
\qquad \epsilon ^{\alpha \beta \gamma \delta \epsilon }\partial _{\alpha
}f_{\beta \gamma }=0  
\label{eqn:16}
\end{equation}
where $e_{0}/\lambda $ is identified as the dimensionless charge $e$, and
the current $j^{\alpha }$ associated with an event $X^{\alpha }=\Bigl(X^{\mu
}(\tau ),\tau \Bigr)$ is given by 
\begin{equation}
j^{\alpha }(x,\tau )=\frac{dX^{\alpha }}{d\tau }\delta ^{4}\Bigl(x^{\mu
}-X^{\mu }(\tau )\Bigr)\ .  
\label{eqn:17}
\end{equation}
At the quantum level, the current is given by (\ref{eqn:6}).

In analogy to the concatenation of the current in (\ref{eqn:8}), we see that
under the boundary conditions $f^{5\mu }\rightarrow 0$, pointwise as $\tau
\rightarrow \pm \infty $, we recover the Maxwell fields from the pre-Maxwell
fields as 
\begin{equation}
\partial _{\nu }F^{\mu \nu }=eJ^{\mu }\qquad \qquad \epsilon ^{\mu
\nu \rho \lambda }\partial _{\mu }F_{\nu \rho }=0  
\label{eqn:18}
\end{equation}
where 
\begin{equation}
F^{\mu \nu }(x)=\int_{-\infty }^{\infty }d\tau \ f^{\mu \nu }(x,\tau )\qquad
\qquad {\rm and}\qquad \qquad A^{\mu }(x)=\int_{-\infty }^{\infty }d\tau \
a^{\mu }(x,\tau )\ .  
\label{eqn:19}
\end{equation}
Since $e_{0}a_{\mu }$ and $eA_{\mu }$ must have the same dimensions, it
follows from (\ref{eqn:19}) that $\lambda $ (and hence $e_{0}=\lambda e$)
must have dimensions of time. Although the parameter $\lambda $ does not
appear in the field equations (\ref{eqn:16}), it does appear in the Lorentz
force (\ref{eqn:12}) through $e_{0}$. The presence of this dimensional
parameter in the equations of motion is a characteristic problem in the
classical formalism.

Although the physical Lorentz covariance of the current $j^{\alpha }$ breaks
the higher symmetry of the free field equations to O(3,1), the wave equation 
\begin{equation}
\partial _{\alpha }\partial ^{\alpha }\ a^{\beta }=(\partial _{\mu }\partial
^{\mu }+\partial _{\tau }\partial ^{\tau })\ a^{\beta }=(\partial _{\mu
}\partial ^{\mu }+\sigma \;\partial _{\tau }^{2})\ a^{\beta }=-\frac{e}{c}\
j^{\beta }\ ,  
\label{eqn:20}
\end{equation}
reflects the causal properties of the higher symmetry through the operator
on the left hand side. The classical Green's function for (\ref{eqn:20}),
defined through 
\begin{equation}
\partial _{\alpha }\partial ^{\alpha }G(x,x^{5})=-\delta ^{4}(x,x^{5})\ ,
\label{eqn:21}
\end{equation}
is given by \cite{green} 
\begin{equation}
G(x,x^{5})=-{\frac{1}{{4\pi }}}\delta (x^{2})\delta (x^{5})-{\frac{1}{{2\pi
^{2}}}}{\frac{\partial }{{\partial {x^{2}}}}}\ {\frac{{\theta (-\sigma
g_{\alpha \beta }x^{\alpha }x^{\beta })}}{{\sqrt{-\sigma g_{\alpha \beta
}x^{\alpha }x^{\beta }}}}}\ .  
\label{eqn:22}
\end{equation}
It follows from (\ref{eqn:20}) and (\ref{eqn:21}) that the potential induced
by a known current is given by 
\begin{equation}
a^{\beta }(x,\tau ) =
-e\int d^{4}x^{\prime }d\tau \ G(x-x^{\prime },\tau -\tau ^{\prime })\
j^{\beta }(x^{\prime },\tau ^{\prime })\ .  
\label{eqn:23}
\end{equation}
Under concatenation, the first term of (\ref{eqn:22}) becomes the Maxwell
Green's function 
\begin{equation}
D(x)=-{\frac{1}{{4\pi }}}\ \delta (x^{2})\ ,  
\label{eqn:24}
\end{equation}
while the second term --- which induces spacelike or timelike correlations 
\cite{green}, depending on the signature $\sigma $ --- vanishes. This
concatenation guarantees that the Maxwell potential is related to the
Maxwell current in the usual manner: 
\begin{eqnarray}
A^{\mu }(x) &=&\int d\tau \ a^{\mu }(x,\tau )  \nonumber \\
&=&-e\int d\tau \int d^{4}x^{\prime }d\tau ^{\prime }\ G(x-x^{\prime },\tau
-\tau ^{\prime })\ j^{\mu }(x^{\prime },\tau ^{\prime })  \nonumber \\
&=&-e\int d^{4}x^{\prime }d\tau ^{\prime }\ \Bigl[\int d\tau G(x-x^{\prime
},\tau -\tau ^{\prime })\Bigr]\ j^{\mu }(x^{\prime },\tau ^{\prime }) 
\nonumber \\
&=&-e\int d^{4}x^{\prime }D(x-x^{\prime })\ J^{\mu }(x^{\prime })\ .
\label{eqn:25}
\end{eqnarray}
Therefore, we will refer to the Maxwell and the correlation terms of the
Green's function and the induced potentials.

The off-shell quantum electrodynamics, associated with the action 
\begin{equation}
{\rm S}=\int d^{4}xd\tau \left\{ \psi ^{*}(i\partial _{\tau
}+e_{0}a_{5})\psi -\frac{1}{2M}\psi ^{*}(-i\partial _{\mu }-e_{0}a_{\mu
})(-i\partial ^{\mu }-e_{0}a^{\mu })\psi -\frac{\lambda }{4}f_{\alpha \beta
}f^{\alpha \beta }\right\} \ ,  
\label{eqn:action}
\end{equation}
has been worked out \cite{qed}. Manifestly covariant quantization has been
given canonically \cite{shnerb,qed} and in path integral form \cite
{jaime,qed}, and the perturbation theory developed \cite{qed}. The Feynman
rules have been used to calculate the scattering cross section for two
identical particles; this cross section reduces to the standard Klein-Gordon
expression when no mass exchange is permitted \cite{qed}, but for non-zero
mass exchange, the forward and reverse poles each split into two and move
away from the 0 and 180 degree directions. The off-shell quantum
electrodynamics is counter-term renormalizable when the photon mass spectrum
is cut off; without the cut-off, the mass integration in the loops cannot be
controlled. We will see below that this cut-off may be introduced through a
the higher order kinetic term introduced below.

\subsection{Classical Coulomb Problem}

The classical Coulomb problem was studied in the framework of the classical
off-shell electromagnetic theory, in \cite{larry}. Posing the classical
equations of motion for an event moving in the field induced by a second
`static' event (an event moving uniformly along the time axis) the
straightforward solution was seen to present a number of difficulties.
First, the events can only interact under very specific circumstances,
leading to a general picture of co-moving, non-interacting charged
particles. Second, under interaction, the particle motion was shown to be
piecewise linear, rather than smooth. The mathematical difficulties derive
from the form of the classical current, which can be seen from (\ref{eqn:17})
to have delta-function support on the worldline of the event. Thus, the
source event 
\begin{equation}
X^{0}(\tau )=\tau \qquad \qquad \vec{X}(\tau )=0  
\label{path}
\end{equation}
induces the current 
\begin{equation}
j^{0}(x,\tau )=j^{5}(x,\tau )=\frac{dX^{0}}{d\tau }\delta (x^{0}-\tau
)\delta ^{3}(\vec{x})=\ \delta (t-\tau )\delta ^{3}(\vec{x})\ ,
\label{path current}
\end{equation}
with $\vec{j}(x,\tau )=0$. Moreover, the leading term in the Green's
function (\ref{eqn:22}) enforces absolute simultaneity in $\tau $, and
reproduces a delta-function in the potential 
\begin{eqnarray}
a^{0}(x,\tau ) &=&a^{5}(x,\tau )=-\frac{e}{4\pi R}\ \frac{1}{2}
%TCIMACRO{\TeXButton{TeX field}{\Bigl} }
%BeginExpansion
\Bigl
%EndExpansion
[\delta (t-R-\tau )+\delta (t+R-\tau )
%TCIMACRO{\TeXButton{TeX field}{\Bigr} }
%BeginExpansion
\Bigr
%EndExpansion
] \\
\vec{a}(x,\tau ) &=&0\ 
\end{eqnarray}
where $R=\left| \vec{x}\right| $. The second term in the Green's function
makes no contribution for non-accelerated sources \cite{larry}. As required
for the Maxwell part, 
\begin{equation}
A^{0}(x)=\int d\tau \ a^{0}(x,\tau )=-\frac{e}{4\pi R}\ ,
\end{equation}
the concatenated potential has the form of the Coulomb potential induced by
a ``fixed'' source. Nevertheless, according to the underlying pre-Maxwell
dynamics, as given in the classical case by the Lorentz force equations 
\begin{eqnarray}
M\ \ddot{x}^{0} &=&e_{0}\ f^{0\alpha }\dot{x}_{\alpha }=-\lambda e\ 
%TCIMACRO{\TeXButton{TeX}{\Bigl} }
%BeginExpansion
\Bigl
%EndExpansion
[(\partial _{k}a^{0})\dot{x}^{k}+\sigma (\partial _{0}+\sigma \partial
_{\tau })\ a^{0}
%TCIMACRO{\TeXButton{TeX}{\Bigr} }
%BeginExpansion
\Bigr
%EndExpansion
] \\
M\ \ddot{x}^{k} &=&e_{0}\ f^{k\alpha }\dot{x}_{\alpha }=-\lambda e\
(\partial _{k}a^{0})(\dot{x}^{0}-\sigma )\ ,
\end{eqnarray}
a test event moving in the field of this `static' event will experience no
interaction unless it crosses through $t\pm R-\tau =0$, and in most cases
the source and test event will appear co-moving and non-interacting in
concatenated laboratory observations. 

Naturally, the interaction condition
depends upon the origin of motion for the source, and a different condition
would apply if the source had been parameterized as $X^{0}(\tau )=\tau
-\varepsilon $. In the classical context, by choosing a 
parameterization for the source and test events, 
$x_{1}^{\mu }\left( \tau \right) $ and $x_{2}^{\mu }\left(\tau \right) $
using a common $\tau$, we implicitly impose
a $\tau $-correlation on the interaction through the $\tau $-simultaneity of
the Green's function. These difficulties do not arise in off-shell quantum
mechanics, because the assumption of sharp asymptotic mass states leads to
complete uncertainty in regard to the location of any event with respect to $
\tau $, and hence to the $\tau $-correlation of a pair of events.

In order to arrive at a reasonable solution to the classical Coulomb
problem, a distribution function was applied {\it ad hoc} to the
field-inducing current in \cite{larry}, in order to smooth out the behavior
of the induced field. It was shown that the smoothed current 
\begin{equation}
j_{\varphi }^{\alpha }(x,\tau )=\int_{-\infty }^{\infty }ds\ \varphi (\tau
-s)\ j^{\alpha }(x,s)  
\label{smoothing}
\end{equation}
in which $\varphi $ is the Laplace distribution 
\begin{equation}
\varphi (\tau )=\frac{1}{2\lambda }e^{-|\tau |/\lambda }
\label{smoothing function}
\end{equation}
leads to a classical Yukawa potential, with interpretation of $\lambda $ as
the mass spectrum of the photons mediating the interaction. Moreover, the extra factor of
$1/\lambda$ combines with $e_0$ in the Lorentz force equations (\ref{eqn:12}) to provide
the correct dimensionless coupling $e$. Since the
smoothing distribution satisfies 
\begin{equation}
\int_{-\infty }^{\infty }d\tau \ \varphi (\tau )=1
\end{equation}
the concatenated Maxwell current is not affected by the integration (\ref
{smoothing}).

The current associated with a `static' particle is now spread out along the
time axis as 
\begin{equation}
j_{\varphi }^{0}(x,\tau )=\int_{-\infty }^{\infty }d\alpha \ \varphi (\alpha
)\ j^{0}(x,\tau -\alpha )=\delta ^{3}(\vec{x})\ \varphi (t-\tau )\ ,
\label{new current}
\end{equation}
as is the pre-Maxwell potential induced by this current 
\begin{equation}
a_{\varphi }^{0}(x,\tau )=-\frac{e}{4\pi R}\ \frac{1}{2}
%TCIMACRO{\TeXButton{TeX field}{\Bigl} }
%BeginExpansion
\Bigl
%EndExpansion
[\varphi (t-R-\tau )+\varphi (t+R-\tau )
%TCIMACRO{\TeXButton{TeX field}{\Bigr} }
%BeginExpansion
\Bigr
%EndExpansion
]\ .
\end{equation}
Under the Laplace distribution, (\ref{smoothing function}) becomes 
\begin{equation}
a_{\varphi }^{0}(x,\tau )=-\frac{e}{4\pi R}\ \frac{1}{2\lambda }\ \frac{1}{2}
\ \left[ e^{-\bigl|t-R-\tau \bigr|/\lambda }+e^{-\bigl|t+R-\tau \bigr|
/\lambda }\right] \ ,  
\label{eqn:76}
\end{equation}
and the motion of the test event under this potential coincides with
classical Coulomb scattering for a small enough limit $\lambda $ on the mass
spectrum of the photons.

Solving the resulting equations of motion in this model, it was shown that
trajectories are indistinguishable from the Maxwell case when $\lambda
>10^{-6}$ seconds, corresponding to a photon mass $m_{\gamma }=1/\lambda
\sim 10^{-9}$ eV. If we take the accepted experimental error in the photon
mass as the actual mass of the photon, then $m_{\gamma }\simeq 6\times
10^{-16}$ eV \cite{pdg}, which corresponds to $\lambda \simeq 1$ second.

Two closely related interpretations of the smoothing process in (\ref{new
current}) were proposed in \cite{larry}. According to the first
interpretation, the smoothing represents an alternative model for the
relationship between events and particles, in which particle currents are
described as a distribution of event currents with different initial
conditions in $\tau $. In the second interpretation, an uncertainty must be
introduced in the mutual $\tau $-correlation of the events. While these
interpretations may hint at some potentially interesting notion of
`classical decoherence' and the Laplace distribution is suggestive of some
Poisson process underlying the formation of particle currents, the mechanism
remains {\it ad hoc} and leaves no clear path to quantization.

\subsection{Nonlocal Electromagnetic Kinetic Term}

In this paper, we return to classical pre-Maxwell electromagnetic theory,
and propose a more general mechanism for smoothing the event current,
involving a modification of the kinetic term for the electromagnetic field.
We discuss the possible alternatives and choose a method which preserves
Lorentz and gauge invariance of the action, and permits first and second
quantization.

\section{A Modified Electromagnetic Action}

The classical electromagnetic action may be taken from (\ref{eqn:14}) to be 
\begin{equation}
S_{em}=\int d^{4}x\,d\tau \,\left[ e_{0}j^{\alpha }a_{\alpha }-\frac{\lambda 
}{4}f^{\alpha \beta }f_{\alpha \beta }\right] \quad ,  
\label{em_action}
\end{equation}
and it would be convenient to make the replacement 
\begin{equation}
j^{\alpha }(x,\tau )\rightarrow j_{\varphi }^{\alpha }(x,\tau
)=\int_{-\infty }^{\infty }ds\ \varphi (\tau -s)\ j^{\alpha }(x,s)
\label{smoothed interaction}
\end{equation}
in (\ref{em_action}), leading directly to modified field equations 
\begin{equation}
\partial _{\beta }f^{\alpha \beta }(x,\tau )=ej_{\varphi }^{\alpha
}\ (x,\tau )\quad ,  
\label{smoothed p-M}
\end{equation}
in which the fields are induced by the smoothed currents. Equation (\ref
{smoothed p-M}) reminds us to check for current conservation explicitly.
Integrating by parts and taking the distribution to be even $\varphi (-\tau
)=\varphi (\tau )$, we indeed find 
\begin{eqnarray}
\partial _{\alpha }\ j_{\varphi }^{\alpha }\left( x,\tau \right) &=&\partial
_{\alpha }\int_{-\infty }^{\infty }ds\ \varphi (\tau -s)\ j^{\alpha }(x,s) 
\nonumber \\
&=&\partial _{\mu }\int_{-\infty }^{\infty }ds\ \varphi (\tau -s)\ j^{\mu
}(x,s)+\partial _{\tau }\int_{-\infty }^{\infty }ds\ \varphi
(\tau -s)\ j^{5}(x,s)  \nonumber \\
&=&\int_{-\infty }^{\infty }ds\ \varphi (\tau -s)\ \partial _{\mu }j^{\mu
}(x,s)-\int_{-\infty }^{\infty }ds\ \partial _{s}\varphi (\tau
-s)\ j^{5}(x,s)  \nonumber \\
&=&\int_{-\infty }^{\infty }ds\ \varphi (\tau -s)\ \partial _{\mu }j^{\mu
}(x,s)+\int_{-\infty }^{\infty }ds\ \varphi (\tau -s)\ \partial
_{s}j^{5}(x,s)  \nonumber \\
&=&\int_{-\infty }^{\infty }ds\ \varphi (\tau -s)\ \left[ \partial _{\mu
}j^{\mu }(x,s)+\partial _{5}j^{5}(x,s)\right]  \nonumber \\
\partial _{\alpha }\ j_{\varphi }^{\alpha }\left( x,\tau \right) &=&0\ .
\label{curr-cons}
\end{eqnarray}
As usual, current conservation guarantees classical gauge invariance under
(\ref{gauge}) as 
\begin{eqnarray}
\int d^{4}x\,d\tau \,a_{\alpha }(x,\tau )\,\ j_{\varphi }^{\alpha }\left(
x,\tau \right) &\rightarrow &\int d^{4}x\,d\tau \,a_{\alpha }(x,\tau )\,\
j_{\varphi }^{\alpha }\left( x,\tau \right)  \nonumber \\
&&+\int d^{4}x\,d\tau \,\partial _{\alpha }\Lambda (x,\tau )\,\,\ j_{\varphi
}^{\alpha }\left( x,\tau \right)  \nonumber \\
\int d^{4}x\,d\tau \,\partial _{\alpha }\Lambda (x,\tau )\,\,\ j_{\varphi
}^{\alpha }\left( x,\tau \right) &=&\int d^{4}x\,d\tau \,\partial _{\alpha
}\left[ \Lambda (x,\tau )\,\,\ j_{\varphi }^{\alpha }\left( x,\tau \right)
\right]  \nonumber \\
&&-\int d^{4}x\,d\tau \,\partial _{\alpha }\Lambda (x,\tau )\,\left[
\,\partial _{\alpha }\ j_{\varphi }^{\alpha }\left( x,\tau \right) \right] 
\nonumber \\
&=&0
\end{eqnarray}

Although the replacement in (\ref{smoothed interaction}) is manifestly
O(3,1) covariant and gauge invariant, it is not clear how it may be applied
in the quantum case (\ref{eqn:action}), when the event current is formed
from bilinear combinations of the field amplitudes. Morover, the underlying
logic of gauge theory requires that we regard the matter amplitudes $\psi
(x,\tau )$ as more fundamental than the gauge fields $a_{\alpha }(x,\tau )$,
and hence the form of the matter part of the action should be respected.

On the other hand, the kinetic term $\frac{\lambda }{4}f^{\alpha \beta
}f_{\alpha \beta }$ for the gauge fields was introduced on essentially
formal grounds --- providing a term with first order derivatives of the
fields, which is a Lorentz scalar, and gauge invariant. Because the field
strengths $f^{\alpha \beta }$ are themselves gauge invariant, a kinetic term
of the form 
\begin{equation}
-\frac{\lambda }{4}\int d^{4}x\,d\tau \,ds\ f^{\alpha \beta }(x,\tau
)\,\,\Phi (\tau -s)\ f_{\alpha \beta }\left( x,s\right)  
\label{new kinetic}
\end{equation}
retains the invariances of (\ref{em_action}), with no assumptions on the
symmetry properties of the distribution $\Phi (\tau )$ --- the symmetic
integration in (\ref{new kinetic}) will vanish over the anti-symmetric part
of the distribution function. The action 
\begin{equation}
S_{em}=\int d^{4}x\,d\tau \,\left[ e_{0}j^{\alpha }(x,\tau )a_{\alpha
}(x,\tau )-\frac{\lambda }{4}\ f^{\alpha \beta }(x,\tau )\,\int ds\,\Phi
(\tau -s)f_{\alpha \beta }\left( x,s\right) \right]  
\label{mod-action}
\end{equation}
leads to modified inhomogeneous pre-Maxwell equations 
\begin{equation}
\partial _{\beta }\int ds\,\Phi (\tau -s)f^{\alpha \beta }(x,s)=e\ j^{\alpha
}(x,\tau )\quad ,
\end{equation}
and by choosing $\Phi (\tau )$ as the inverse of $\varphi (\tau )$, such
that 
\begin{equation}
\int_{-\infty }^{\infty }ds\ \Phi (\tau -s)\,\varphi (s-r)=\delta (\tau
-r)\quad ,  
\label{inv_pm}
\end{equation}
it is possible to invert (\ref{inv_pm}) to obtain the desired result 
\begin{eqnarray}
\partial _{\beta }\int d\tau ^{\prime }\ \varphi (\tau -\tau ^{\prime })\int
ds\,\Phi (\tau ^{\prime }-s)f^{\alpha \beta }(x,s) &=&e\int d\tau ^{\prime
}\ \varphi (\tau -\tau ^{\prime })\ j^{\alpha }(x,\tau ^{\prime }) \\
\partial _{\beta }\int ds\,\delta (\tau -s)f^{\alpha \beta }(x,s) &=&e\int
ds\ \varphi (\tau -s)\ j^{\alpha }(x,s) \\
\partial _{\beta }f^{\alpha \beta }(x,\tau ) &=&e\,j_{\varphi }^{\alpha }\
(x,\tau )\,\;\;.
\end{eqnarray}
For the Laplace distribution (\ref{smoothing function}) used in \cite{larry}
, the inverse function may be found readily in the ``mass domain''.
Expressing the distribution as a Fourier integral over ``conjugate mass'', 
\begin{equation}
\varphi (\tau )=\frac{1}{2\lambda }e^{-|\tau |/\lambda }=\int \frac{d\kappa 
}{2\pi }\,\widetilde{\varphi }\left( \kappa \right) \,e^{-i\kappa \tau }
\end{equation}
we find the mass regulation kernel 
\begin{equation}
\widetilde{\varphi }\left( \kappa \right) =\int d\tau \,e^{i\kappa \tau
}\,\varphi (\tau )=\frac{1}{2\lambda }\int d\tau \,e^{i\kappa \tau
}\,e^{-|\tau |/\lambda }=\frac{1}{1+\left( \lambda \kappa \right) ^{2}}\,\,.
\end{equation}
Similarly expressing the inverse function as a Fourier integral, 
\begin{equation}
\Phi (\tau )=\int \frac{d\kappa }{2\pi }\,\widetilde{\Phi }\left( \kappa
\right) \,e^{-i\kappa \tau }
\end{equation}
it follows that 
\begin{equation}
\int_{-\infty }^{\infty }ds\ \Phi (\tau -s)\,\varphi (s-r)=\delta (\tau
-r)\qquad \Rightarrow \qquad \widetilde{\Phi }\left( \kappa \right) =\frac{1
}{\widetilde{\varphi }\left( \kappa \right) }=1+\left( \lambda \kappa
\right) ^{2}
\end{equation}
and so 
\begin{eqnarray}
\Phi (\tau ) &=&\int \frac{d\kappa }{2\pi }\,\left[ 1+\left( \lambda \kappa
\right) ^{2}\right] \,e^{-i\kappa \tau }  \nonumber \\
&=&\int \frac{d\kappa }{2\pi }\,e^{-i\kappa \tau }+\lambda ^{2}\int \frac{
d\kappa }{2\pi }\kappa ^{2}e^{-i\kappa \tau }  \nonumber \\
&=&\int \frac{d\kappa }{2\pi }\,e^{-i\kappa \tau }+\lambda ^{2}\int \frac{
d\kappa }{2\pi }\left( i\partial _{\tau }\right) ^{2}\,e^{-i\kappa \tau } 
\nonumber \\
&=&\int \frac{d\kappa }{2\pi }\,e^{-i\kappa \tau }+\lambda ^{2}\left( i\frac{
d}{d\tau }\right) ^{2}\int \frac{d\kappa }{2\pi }\,e^{-i\kappa \tau } 
\nonumber \\
&=&\delta \left( \tau \right) -\lambda ^{2}\delta ^{\prime \prime }\left(
\tau \right) \,\ .  
\label{inverse function}
\end{eqnarray}
Inserting (\ref{inverse function}) into (\ref{new kinetic}), the modified
kinetic term for the gauge fields is 
\begin{eqnarray}
S_{em-kinetic} &=&\frac{\lambda }{4}\int d^{4}x\,d\tau \,ds\ f^{\alpha \beta
}(x,\tau )\,\,\Phi (\tau -s)\ f_{\alpha \beta }\left( x,s\right)  \nonumber
\\
&=&\frac{\lambda }{4}\int d^{4}x\,d\tau \,ds\ f^{\alpha \beta }(x,\tau
)\,\,\left[ \delta \left( \tau -s\right) -\lambda ^{2}\delta ^{\prime \prime
}\left( \tau -s\right) \right] \ f_{\alpha \beta }\left( x,s\right) 
\nonumber \\
&=&\frac{\lambda }{4}\int d^{4}x\,d\tau \,\ f^{\alpha \beta }(x,\tau )\,\
f_{\alpha \beta }\left( x,\tau \right) -\frac{\lambda ^{3}}{4}\int
d^{4}x\,d\tau \,\ f^{\alpha \beta }(x,\tau )\,\,\partial _{\tau }^{2}\
f_{\alpha \beta }\left( x,\tau \right)  \nonumber \\
&=&\frac{\lambda }{4}\int d^{4}x\,d\tau \,\ f^{\alpha \beta }(x,\tau )\,\
f_{\alpha \beta }\left( x,\tau \right) +\frac{\lambda ^{3}}{4}\int
d^{4}x\,d\tau \,\ 
%TCIMACRO{\TeXButton{big}{\Bigl} }
%BeginExpansion
\Bigl
%EndExpansion
(\partial _{\tau }f^{\alpha \beta }(x,\tau )
%TCIMACRO{\TeXButton{big}{\Bigr} }
%BeginExpansion
\Bigr
%EndExpansion
)
%TCIMACRO{\TeXButton{big}{\Bigl} }
%BeginExpansion
\Bigl
%EndExpansion
(\partial _{\tau }\ f_{\alpha \beta }\left( x,\tau \right) \,
%TCIMACRO{\TeXButton{big}{\Bigr} }
%BeginExpansion
\Bigr
%EndExpansion
)  \nonumber \\
&=&S_{em-kinetic}^{0}+\frac{\lambda ^{3}}{4}\int d^{4}x\,d\tau \,\ 
%TCIMACRO{\TeXButton{big}{\Bigl} }
%BeginExpansion
\Bigl
%EndExpansion
(\partial _{\tau }f^{\alpha \beta }(x,\tau )
%TCIMACRO{\TeXButton{big}{\Bigr} }
%BeginExpansion
\Bigr
%EndExpansion
)
%TCIMACRO{\TeXButton{big}{\Bigl} }
%BeginExpansion
\Bigl
%EndExpansion
(\partial _{\tau }\ f_{\alpha \beta }\left( x,\tau \right) \,
%TCIMACRO{\TeXButton{big}{\Bigr} }
%BeginExpansion
\Bigr
%EndExpansion
)\,\ .  
\label{high_order kin}
\end{eqnarray}
In this form, the modified kinetic term appears as the usual term plus a
higher-order derivative of the fields. Although the higher-order derivative
breaks the formal O(3,2) or O(4,1) symmetry to O(3,1), both Poincar\'{e}
invariance and gauge invariance are preserved.
\newpage
\section{Quantization of the Interacting Theory}

The modified action can be written in the form 
\begin{eqnarray}
{\rm S} &=&\int d^{4}xd\tau \left\{ \psi ^{*}(i\partial _{\tau
}+e_{0}a_{5})\psi -\frac{1}{2M}\psi ^{*}(-i\partial _{\mu }-e_{0}a_{\mu
})(-i\partial ^{\mu }-e_{0}a^{\mu })\psi \right\}  \nonumber \\
&&\quad \quad \quad \quad \quad \quad \quad \quad \quad \quad \quad -\frac{
\lambda }{4}\int d^{4}x\,d\tau \,ds\ f^{\alpha \beta }(x,\tau )\,\,\Phi
(\tau -s)\ f_{\alpha \beta }\left( x,s\right) \,\,,  
\label{new action}
\end{eqnarray}
and we adopt the notation 
\begin{equation}
\ \left( f^{\Phi }\right) _{\alpha \beta }\left( x,\tau \right) =\int ds\
\Phi (\tau -s)\ f_{\alpha \beta }\left( x,s\right) \,\,.  
\label{f-phi}
\end{equation}
As shown in \cite{qed} and \cite{letter}, a straight forward quantization is
obtained by generalizing the Jackiw-Fadeev ``first order quantization''
method \cite{jackiw} to the path integral. Since the canonical structure of
the matter field guarantees first-order form with respect to $\tau $
-derivatives, only the gauge field must be modified. Expanding the kinetic
term 
\begin{eqnarray}
f^{\alpha \beta }\left( f^{\Phi }\right) _{\alpha \beta } &=&f^{\mu \nu
}\left( f^{\Phi }\right) _{\mu \nu }+2f^{5\mu }\left( f^{\Phi }\right)
_{5\mu }  \nonumber \\
&=&f^{\mu \nu }\left( f^{\Phi }\right) _{\mu \nu }-2f^{5\mu }\left( f^{\Phi
}\right) ^{5}\,_{\mu }  \nonumber \\
&=&f^{\mu \nu }\left( f^{\Phi }\right) _{\mu \nu }+2\left[ 2(\partial _{\tau
}a^{\mu }+\partial ^{\mu }a^{5})\left( f^{\Phi }\right) ^{5}\,_{\mu
}-f^{5\mu }\left( f^{\Phi }\right) ^{5}\,_{\mu }\right] \ ,  
\label{eqn:2.10}
\end{eqnarray}
integrating by parts 
\begin{equation}
(\partial _{\tau }a^{\mu }+\partial ^{\mu }a^{5})\left( f^{\Phi }\right)
^{5}\,_{\mu }=(\partial _{\tau }a^{\mu })\left( f^{\Phi }\right) ^{5}\,_{\mu
}-a^{5}\partial ^{\mu }\left( f^{\Phi }\right) ^{5}\,_{\mu }+{\rm divergence}
\ ,  
\label{eqn:2.12}
\end{equation}
and introducing the definitions 
\begin{equation}
\epsilon ^{\mu }=f^{5\mu }\qquad \left( \epsilon ^{\Phi }\right) ^{\mu
}=\left( f^{\Phi }\right) ^{5\mu }\ \ ,  
\label{eqn:2.11}
\end{equation}
the action becomes 
\begin{eqnarray}
{\rm S} &=&\int d^{4}x\,d\tau \,\,\left[ i\psi ^{*}\dot{\psi}-\lambda \dot{a}
^{\mu }\left( \epsilon ^{\Phi }\right) _{\mu }-\frac{1}{2M}\psi
^{*}(-i\partial _{\mu }-e_{0}a_{\mu })(-i\partial ^{\mu }-e_{0}a^{\mu })\psi
\right.  \nonumber \\
&&\left. -\frac{\lambda }{4}f^{\mu \nu }\left( f^{\Phi }\right) _{\mu \nu }-
\frac{\lambda }{2}\epsilon ^{\mu }\left( \epsilon ^{\Phi }\right) _{\mu
}+a_{5}\left( e_{0}\psi ^{*}\psi -\lambda \partial ^{\mu }\left( \epsilon
^{\Phi }\right) _{\mu }\right) \right]  
\label{act1}
\end{eqnarray}
where $\dot{\psi}=\partial _{\tau }\psi $ and $\dot{a}^{\mu }=\partial
_{\tau }a^{\mu }$. Placing the action (\ref{act1}) into a path integral 
\begin{equation}
{\cal Z}=\frac{1}{{\cal N}}\int {\cal D}\psi ^{*}\ {\cal D}\psi \ {\cal D}
a_{\mu }\ {\cal D}a_{5}\ {\cal D}\epsilon _{\mu }e^{iS}  
\label{eqn:3.1}
\end{equation}
and noticing the absence of $\dot{a}^{5}$, the integration over ${\cal D}
a_{5}$ simply places 
\begin{equation}
\delta \left( e_{0}\psi ^{*}\psi -\lambda \partial ^{\mu }\left( \epsilon
^{\Phi }\right) _{\mu }\right)  
\label{constr}
\end{equation}
into the measure. The constraint (\ref{constr}) imposes the Gauss Law in the
pre-Maxwell theory, 
\begin{equation}
\quad \partial ^{\mu }\left( \epsilon ^{\Phi }\right) _{\mu }=\frac{e_{0}}{
\lambda }\psi ^{*}\psi =ej^{5}  
\label{gauss}
\end{equation}
and following Jackiw-Fadeev, we eliminate the constraint by solving (\ref
{gauss}). The solution may be found by a covariant decomposition into
transverse and longitudinal parts 
\begin{equation}
\left( \epsilon ^{\Phi }\right) ^{\mu }=\left( \epsilon _{\perp }^{\Phi
}\right) ^{\mu }+e\partial ^{\mu }[Gj^{5}]  
\label{eqn:2.15}
\end{equation}
where the transverse part satisifies 
\begin{equation}
\partial _{\mu }\left( \epsilon _{\perp }^{\Phi }\right) ^{\mu }=0
\end{equation}
and the longitudinal part is found from 
\begin{equation}
\lbrack Gj^{5}](x,\tau )=\int d^{4}y\ G(x-y)\ j^{5}(y,\tau )
\label{eqn:2.17}
\end{equation}
with the Green's function 
\begin{equation}
G(x-y)=\delta \left( (x-y)^{2}\right) \quad \Rightarrow \quad \Box G=1\ .
\label{eqn:2.18}
\end{equation}
Similarly, we impose the gauge fixing 
\begin{equation}
\delta \left( \partial _{\mu }a^{\mu }-\Lambda \right)
\end{equation}
which we solve through the decomposition 
\begin{equation}
a^{\mu }=a_{\perp }^{\mu }+\partial ^{\mu }[G\Lambda ]\qquad \partial _{\mu
}a_{\perp }^{\mu }=0\ ,
\end{equation}
which further entails 
\begin{equation}
\dot{a}^{\mu }=\dot{a}_{\perp }^{\mu }+\partial ^{\mu }[G\dot{\Lambda}]
\label{eqn:2.21}
\end{equation}
\begin{equation}
f^{\mu \nu }=\partial ^{\mu }a_{\perp }^{\nu }-\partial ^{\nu }a_{\perp
}^{\mu }=f_{\perp {}}^{\mu \nu }  
\label{eqn:2.22}
\end{equation}
\begin{equation}
-i\partial ^{\mu }-e_{0}a^{\mu }=-i\partial ^{\mu }-e_{0}a_{\perp }^{\mu
}-e_{0}\partial ^{\mu }[G\dot{\Lambda}]  
\label{eqn:2.23}
\end{equation}
\begin{equation}
\dot{a}^{\mu }\left( \epsilon ^{\Phi }\right) _{\mu }=\dot{a}_{\perp }^{\mu
}\left( \epsilon _{\perp }^{\Phi }\right) _{\mu }-e\rho [G\dot{\Lambda}]\ \ .
\label{eqn:2.25}
\end{equation}
Having eliminated the transverse components, the path integral is now 
\begin{equation}
{\cal Z}=\frac{1}{{\cal N}}\int {\cal D}\psi ^{*}\ {\cal D}\psi \ {\cal D}
(a_{\perp })_{\mu }\ {\cal D}(\epsilon _{\perp })_{\mu }\ e^{iS}
\label{eqn:3.2}
\end{equation}
where 
\begin{eqnarray}
{\rm S} &=&\int d^{4}x\,d\tau \,\,\left[ i\psi ^{*}\dot{\psi}-\lambda \dot{a}
_{\perp }^{\mu }{}\left( \epsilon _{\perp }^{\Phi }\right) _{\mu }+\lambda
ej^{5}[G\dot{\Lambda}]\right.  \nonumber \\
&&-\frac{1}{2M}\psi ^{*}(-i\partial _{\mu }-e_{0}a_{\perp }{}_{\mu
}-e_{0}\partial _{\mu }[G\dot{\Lambda}])(-i\partial ^{\mu }-e_{0}a{}_{\perp
}^{\mu }-e_{0}\partial ^{\mu }[G\dot{\Lambda}])\psi  \nonumber \\
&&\left. -\frac{\lambda }{4}f{}_{\perp }^{\mu \nu }\left( f_{\perp }^{\Phi
}\right) _{\mu \nu }-\frac{\lambda }{2}\epsilon {}_{\perp }^{\mu }\left(
\epsilon _{\perp }^{\Phi }\right) _{\mu }+\frac{\lambda }{2}
e^{2}j^{5}[Gj^{5}]\ \right] \ \ .  
\label{eqn:2.26}
\end{eqnarray}
The next step is to perform the Gaussian integration over ${\cal D}(\epsilon
_{\perp })_{\mu }$ 
\begin{eqnarray}
\int {\cal D}\epsilon _{\perp }{}_{\mu } &&\!\!\!\!\!\!\!\!\!\!\!\!\!\exp
\left[ -i\int d^{4}x\,d\tau \,\,\,\frac{\lambda }{2}\epsilon {}_{\perp
}^{\mu }\left( \epsilon _{\perp }^{\Phi }\right) _{\mu }+\lambda \dot{a}
_{\perp }^{\mu }{}\left( \epsilon _{\perp }^{\Phi }\right) _{\mu }\right] = 
\nonumber \\
&=&\int {\cal D}\epsilon _{\perp }{}_{\mu }\exp \left[ -i\int d^{4}x\,d\tau
\,\,ds\,\frac{\lambda }{2}\epsilon {}_{\perp }^{\mu }\Phi \epsilon _{\perp
\mu }+\lambda \dot{a}_{\perp }^{\mu }{}\Phi \epsilon _{\perp \mu }\right] 
\nonumber \\
&=&\exp \left[ -\,i\frac{\lambda }{2}\int d^{4}x\,d\tau \,\,ds\ \dot{a}
_{\perp }^{\mu }\Phi \dot{a}_{\perp \mu }\right]  \nonumber \\
&=&\exp \left[ -\,i\frac{\lambda }{2}\int d^{4}x\,d\tau \ \,\dot{a}_{\perp
}^{\mu }\left( \dot{a}_{\perp }^{\Phi }\right) _{\mu }\right] \ \ .
\label{gaussian}
\end{eqnarray}
The path integral now takes the form 
\begin{equation}
{\cal Z}=\frac{1}{{\cal N}}\int {\cal D}\psi ^{*}\ {\cal D}\psi \ {\cal D}
(a_{\perp })_{\mu }\ e^{iS}
\end{equation}
where 
\begin{eqnarray}
{\rm S} &=&\int d^{4}x\,d\tau \,\,\left[ i\psi ^{*}\dot{\psi}+\lambda
ej^{5}[G\dot{\Lambda}]\right.  \nonumber \\
&&-\frac{1}{2M}\psi ^{*}(-i\partial _{\mu }-e_{0}a_{\perp }{}_{\mu
}-e_{0}\partial _{\mu }[G\dot{\Lambda}])(-i\partial ^{\mu }-e_{0}a{}_{\perp
}^{\mu }-e_{0}\partial ^{\mu }[G\dot{\Lambda}])\psi  \nonumber \\
&&\left. -\frac{\lambda }{4}f{}_{\perp }^{\mu \nu }\left( f_{\perp }^{\Phi
}\right) _{\mu \nu }+\frac{\lambda }{2}\dot{a}_{\perp }^{\mu }\left( \dot{a}
_{\perp }^{\Phi }\right) _{\mu }+\frac{\lambda }{2}e^{2}j^{5}[Gj^{5}]\
\right] \ \ ,
\end{eqnarray}
and eliminating the gauge fixing with the transformation 
\begin{equation}
\psi \longrightarrow e^{ie_{0}[G\Lambda ]}\psi  
\label{eqn:2.27}
\end{equation}
we are left with 
\begin{equation}
{\cal Z}=\frac{1}{{\cal N}}\int {\cal D}\psi ^{*}\ {\cal D}\psi \ {\cal D}
(a_{\perp })_{\mu }\ e^{iS}  
\label{eqn:3.4}
\end{equation}
where the action has the unconstrained transverse form 
\begin{eqnarray}
{\rm S} &=&\int d^{4}xd\tau \left\{ i\psi ^{*}\dot{\psi}-\frac{1}{2M}\psi
^{*}(-i\partial _{\mu }-e_{0}(a_{\perp })_{\mu })(-i\partial ^{\mu
}-e_{0}(a_{\perp })^{\mu })\psi \right.  \nonumber \\
&&\left. -\frac{\lambda }{4}(f_{\perp })^{\mu \nu }\left( f_{\perp }^{\Phi
}\right) _{\mu \nu }+\frac{\lambda }{2}(\dot{a}_{\perp })_{\mu }(\dot{a}
_{\perp }^{\Phi })^{\mu }+\frac{\lambda }{2}e^{2}j^{5}[Gj^{5}]\right\} \ \ .
\label{eqn:3.5}
\end{eqnarray}
Finally, making the rearrangements 
\begin{eqnarray}
-\frac{\lambda }{4}(f_{\perp })^{\mu \nu }\left( f_{\perp }^{\Phi }\right)
_{\mu \nu } &=&-\frac{\lambda }{4}[\partial ^{\mu }a_{\perp }^{\nu
}-\partial ^{\nu }a_{\perp }^{\mu }][\partial _{\mu }\left( a_{\perp }^{\Phi
}\right) _{\nu }-\partial _{\nu }\left( a_{\perp }^{\Phi }\right) _{\mu }] 
\nonumber \\
&=&-\frac{\lambda }{2}\left[ \left( \partial ^{\mu }a_{\perp }^{\nu }\right)
\left( \partial _{\mu }\left( a_{\perp }^{\Phi }\right) _{\nu }\right)
-\left( \partial ^{\nu }a^{\mu }\right) \left( \partial _{\mu }\left(
a_{\perp }^{\Phi }\right) _{\nu }\right) \right]  \nonumber \\
&=&\frac{\lambda }{2}a_{\mu }[g^{\mu \nu }\Box -\partial ^{\mu }\partial
^{\nu }]\left( a_{\perp }^{\Phi }\right) _{\nu }+{\rm divergence}
\label{eqn:3.6}
\end{eqnarray}
and 
\begin{equation}
(\dot{a}_{\perp })_{\mu }(\dot{a}_{\perp }^{\Phi })^{\mu }=-(a_{\perp
})_{\mu }[g^{\mu \nu }\partial _{\tau }^{2}](a_{\perp }^{\Phi })_{\nu }+{\rm 
divergence}  
\label{eqn:3.7}
\end{equation}
the action becomes 
\begin{eqnarray}
{\rm S} &=&\int d^{4}xd\tau \Bigl\{i\psi ^{*}\dot{\psi}-\frac{1}{2M}\psi
^{*}(-i\partial _{\mu }-e_{0}(a_{\perp })_{\mu })(-i\partial ^{\mu
}-e_{0}(a_{\perp })^{\mu })\psi  \nonumber \\
&&+\frac{\lambda }{2}(a_{\perp })_{\mu }[\Box +\sigma \partial _{\tau
}^{2}](a_{\perp }^{\Phi })^{\mu }+\frac{\lambda }{2}e^{2}j^{5}[Gj^{5}]\Bigr
\}\ \ .  
\label{eqn:3.9}
\end{eqnarray}
Inserting the explicit form for the inverse Laplace distribution $\Phi (\tau
)$ 
\begin{eqnarray}
{\rm S} &=&\int d^{4}xd\tau \Bigl\{i\psi ^{*}\dot{\psi}-\frac{1}{2M}\psi
^{*}(-i\partial _{\mu }-e_{0}(a_{\perp })_{\mu })(-i\partial ^{\mu
}-e_{0}(a_{\perp })^{\mu })\psi  \nonumber \\
&&+\frac{\lambda }{2}e^{2}j^{5}[Gj^{5}]\Bigr\}+\int d^{4}xd\tau ds\frac{
\lambda }{2}(a_{\perp })^{\mu }[\Box +\sigma \partial _{\tau }^{2}]\Phi
{}a_{\perp \mu }
\end{eqnarray}
\begin{eqnarray}
{\rm S} &=&\int d^{4}xd\tau \Bigl\{i\psi ^{*}\dot{\psi}-\frac{1}{2M}\psi
^{*}(-i\partial _{\mu }-e_{0}(a_{\perp })_{\mu })(-i\partial ^{\mu
}-e_{0}(a_{\perp })^{\mu })\psi  \nonumber \\
&&+\frac{\lambda }{2}(a_{\perp })^{\mu }[\Box +\sigma \partial _{\tau
}^{2}]\left( 1-\lambda ^{2}\partial _{\tau }^{2}\right) {}a_{\perp \mu }+
\frac{\lambda }{2}e^{2}j^{5}[Gj^{5}]\Bigr\}\ \ .  
\label{act3}
\end{eqnarray}
we can now read the Feynman rules from the action (\ref{act3}). We may
summarize the Feynman rules for the momentum space Green's functions as
follows:

\begin{enumerate}
\item  For each matter field propagator, draw a directed line associated
with the factor 
\begin{equation}
\frac{1}{(2\pi )^{5}}\frac{-i}{\frac{1}{2M}p^{2}-P-i\epsilon }
\label{mat-prop}
\end{equation}

\item  For each photon propagator, draw a photon line associated with the
factor 
\begin{equation}
\frac{1}{\lambda }\left[ g^{\mu \nu }-\frac{k^{\mu }k^{\nu }}{k^{2}}\right] 
\frac{-i}{k^{2}+\kappa ^{2}-i\epsilon }\ \ \frac{1}{1+\lambda ^{2}\kappa ^{2}
}  
\label{phot-prop}
\end{equation}

\item  For the three-particle interaction, write the vertex factor 
\begin{equation}
\frac{e_{0}}{2M}i(p+p^{\prime })^{\nu }\ (2\pi )^{5}\delta ^{4}(p-p^{\prime
}-k)\delta (P-P^{\prime }-\kappa )  
\label{vert1}
\end{equation}

\item  For the four-particle interaction, write the vertex factor 
\begin{equation}
\mbox{\qquad}\frac{-ie_{0}^{2}}{M}(2\pi )^{5}g_{\mu \nu }\delta
^{4}(k-k^{\prime }-p^{\prime }+p)\delta (-\kappa +\kappa ^{\prime
}+P^{\prime }-P)  
\label{vert2}
\end{equation}
\end{enumerate}

These Feynman rules are identical to those given in \cite{qed} and \cite
{letter}, with the addition of the mass regulating factor $\frac{1}{
1+\lambda ^{2}\kappa ^{2}}$ in (\ref{phot-prop}). It was shown in \cite{qed}
that with an appropriate mass cut-off, the off-shell quantum electrodynamics
can be counter-term renormalized. In fact, since the matter propagator is
retarded, in the sense that 
\begin{equation}
G(x-x^{\prime },\tau -\tau ^{\prime })=0,\qquad \tau <\tau ^{\prime }\quad
,
\end{equation}
there are no matter loops in the resulting theory, and the mass-regulated theory
is super-renomalizable. With the mass regulation offered by the higher-order
kinetic term, pre-Maxwell quantum electrodynamics is essentially finite ---
only the one-photon loop self-energy diagram diverges as 
\begin{equation}
G_{0}^{(2)}(p)\ \left( (2\pi )^{5}\frac{ie_{0}^{2}}{M}\right) ^{2}\frac{1}{
\lambda }\int d^{4}qdQ\frac{-i}{q^{2}+Q^{2}-i\epsilon }\ \ \frac{1}{
1+\lambda ^{2}Q^{2}} \ G_{0}^{(2)}(p)
\end{equation}
and this diagram is renormalized by shifting the term $i\psi ^{*}\partial
_{\tau }\psi $. The problematic divergent diagram in the unregulated theory
is given by the overlapping two-photon diagram, and contributes the
following term to the matter propagator:

\begin{eqnarray}
&&G_{0}^{(2)}(p)\ G_{0}^{(2)}(p)\left( (2\pi )^{5}\frac{ie_{0}}{2M}\right)
^{4}16p^{4}\int d^{4}qdQd^{4}q^{\prime }dQ^{\prime }  \nonumber \\
&&\mbox{\qquad}\frac{1}{\lambda ^{2}}\frac{1}{q^{2}-Q^{2}}\ \frac{1}{
q^{\prime 2}-Q^{\prime 2}}\ \frac{2M}{(p-q)^{2}-2M(P-Q)}  \nonumber \\
&&\mbox{\qquad}\frac{2M}{(p-q^{\prime })^{2}-2M(P-Q^{\prime })}\ \frac{2M}{
(p-q-q^{\prime })^{2}-2M[P-(Q+Q^{\prime })]}\quad .
\end{eqnarray}
Since this term includes the factor $p^{4}$, it cannot be renormalized by a
counter-term, and a mass cut-off is required. With the regulation provided
by the higher-order kinetic term, this contribution is now completely
finite: 
\begin{eqnarray}
&&G_{0}^{(2)}(p)\ G_{0}^{(2)}(p)\left( (2\pi )^{5}\frac{ie_{0}}{2M}\right)
^{4}16p^{4}\int d^{4}qdQd^{4}q^{\prime }dQ^{\prime }  \nonumber \\
&&\mbox{\qquad}\frac{1}{\lambda ^{2}}\frac{1}{q^{2}-Q^{2}}\ \left\{ \frac{1}{
1-\lambda ^{2}Q^{2}}\right\} \frac{1}{q^{\prime 2}-Q^{\prime 2}}\left\{ 
\frac{1}{1-\lambda ^{2}Q^{\prime 2}}\right\} \ \frac{2M}{(p-q)^{2}-2M(P-Q)} 
\nonumber \\
&&\mbox{\qquad}\frac{2M}{(p-q^{\prime })^{2}-2M(P-Q^{\prime })}\ \frac{2M}{
(p-q-q^{\prime })^{2}-2M[P-(Q+Q^{\prime })]}\quad .
\end{eqnarray}

\section{Conclusion}

We have shown that the addition of the higher-order kinetic term
for the gauge fields provides a Lorentz-covariant and gauge
invariant approach to both mass regulation in pre-Maxwell quantum
electrodynamics, and to a corresponding smoothing of the
classical electromagnetic potential, leading to a reasonable
description of Coulomb scattering.  It is particularly
interesting that this modification of the kinetic term for the
gauge fields is equivalent to the introduction of
parameterization uncertainty into the particle-field interaction.
Moreover, the appearance of the Laplace distribution --- hinting
at an underlying Poisson process assigning events along particle
world lines --- suggests further directions of exploration in
pre-Maxwell theory and higher-order Lagrangians.

%
%%%%%%%%%%%%%%%%%%%%%%%% REFERENCES %%%%%%%%%%%%%%%%%%%%%%%%%%%%%
%

%

%

\end{document}